\documentclass[aps,twocolumn,showrefs,amsmath,amssymb,prx,colors,longbiblography]{revtex4-1} 
 
\usepackage{amsmath}    
\usepackage{amssymb}
\usepackage{graphicx} 
\usepackage{hyperref} 
\usepackage[english]{babel}
\usepackage{tabularx}
\usepackage{bm}
\usepackage{calrsfs}
\usepackage{csquotes}
\usepackage{color} 

\usepackage{xspace}

\usepackage{subfigure}

\def\be{\begin{equation}}     
\def\ee{\end{equation}}
\def\ba{\begin{eqnarray}}
\def\ea{\end{eqnarray}}
\def\<{\left<}
\def\>{\right>} 
\def\({\left(}
\def\){\right)} 
\def\[{\left[}
\def\]{\right]}

\begin{document}

\title{On the thermopower of ionic conductors and ionic capacitors}
\author{Alois W\"urger}
\affiliation{Universit\'e de Bordeaux \& CNRS, LOMA (UMR 5798), 33405 Talence, France}

\begin{abstract}
We theoretically study the thermoelectric response of ionic conductors to an applied temperature gradient. As a main result we find that open and closed systems with respect to charge exchange, result in different expressions for the thermopower which may even take opposite signs. For the experimentally most relevant zero-current steady state, we show that the thermopower of ionic conductors does not depend on the mobilities, contrary to what is known for metals and semiconductors. The different behavior of ionic and electronic conductors is traced back to the unlike conservation laws for ionic carriers and electron-hole pairs. 
\end{abstract} 

\maketitle
 

\section{Introduction}

Thermoelectric materials are extensively studied for energy applications such as the conversion of low-grade waste heat into electrical power.  As an important performance parameter, the thermopower (or Seebeck coefficient) $S$ describes the electric current generated by a temperature gradient, or the voltage difference arising between hot and cold boundaries \cite{Bell2008}.

Onsager's reciprocal relations link the Seebeck coefficient  $S=Q/qT$ to the Peltier heat of transport $Q$, which is the enthalpy carried by a charge $q$ moving in an electric field. Thermoelectric effects were first observed for metals, with $S$ much smaller than the natural unit $k_B/e= 86\mathrm{\mu V/K}$.  Stronger effects occur in semiconductors, where the thermopower ranges from 1 to  20 $k_B/e$ \cite{Bubnova2011,Russ2016,Gregory2018,Xiao2017}. These numbers are rationalized in terms of Mott's formula, accounting for electronic band structure effects, doping, Anderson localization well below the Fermi surface, and an energy-dependent mobility $\mu(E)$ \cite{Cutler1969,Fritsche1971,Kang2017}. 

Ionic conductors differ from electronic devices in two fundamental aspects: First, ions cannot be transferred to electrodes, and thus cannot directly generate thermoelectric currents. Second, in general there are several carrier species and, at least in the absence of redox reactions, the number of each of them is conserved. Accordingly, $S$ consists of the sum of ion-specific contributions, which are independent of concentrations as long as correlation effects are negligible \cite{Onsager1927,Demery2016}. In this paper we consider a binary electrolyte, where the Seebeck coefficient reads as
  \be
  S = \frac{w_+ Q_+ - w_- Q_-}{eT}, 
  \ee  
with weight factors $w_\pm$ and the heats of transport  $Q_\pm$ of cations and anions. In aqueous  solution, the solvation enthalpies of common salt ions are of the order $Q_\pm\sim k_BT$ \cite{Podszus1908,Takeyama1983,Agar1989,Salez2017,Sehnem2020}. The resulting Seebeck coefficent, $S\sim k_B/e$, was shown to drive colloidal thermophoresis \cite{Putnam2005,Wuerger2008,Vigolo2010,Eslahian2014,Lin2018}. In recent years, much higher values up to 300 $k_B/e$ were reported for polymer-based electrolytes \cite{Zhao2016,Wang2017,Li2019}, small mobile ions in gels or solid matrices \cite{Kim2016,Han2020}, or ionic liquids \cite{Zhao2019}. 

There seems to be no general agreement regarding the weight factors $w_\pm$ for the carrier specific contributions to the Seebeck coefficient. If all previous works agree on the fact that they  are proportional to concentration $w_\pm\propto n_\pm$, discrepancies arise with respect to the dependencies on ion valency and mobility; the latter 	appears when identifying $w_\pm$ with Hittorf transport numbers $t_\pm$ that account for the relative conductivity of each ion species.  At present it is not clear which description is correct for the Seebeck coefficient of ionic conductors. This question is of practical interest: Because of the large mobility contrast of polymer electrolytes, different weight factors may even result in opposite signs of the Seebeck coefficient $S$. 

In the present note we study the thermoelectric properties of ionic conductors as open or closed systems, where the former exchange charges with the environment and the latter don't. Starting from two experimental situations with well-defined boundary conditions, we find that the corresponding Seebeck coefficients may significantly differ from each other. We discuss our results in view of recent experiments \cite{Zhao2016,Wang2017,Li2019,Kim2016,Han2020,Zhao2019}, and compare with what is known for electronic materials. 

\section{Ionic conductors}

We consider an electrolyte solution of positive and negative charge carriers with concentrations $n_\pm$  and mobilities $\mu_\pm$. In simple monovalent electrolytes, overall charge neutrality imposes $n_+=n_-$. In complex systems, the concentrations of positive and negative carriers need not to be identical. Then the charge density reads as 
  \be
  \varrho = e(n_+ - n_-) + \varrho_f,  
  \label{eq:4}
  \ee
where $\varrho_f$ are fixed charges. As examples we note polyelectrolyte complexes where the number difference of mobile ions $n_+\neq n_-$ is compensated by the charge  $\varrho_f$ of the  solid matrix \cite{De2017,Ostendorf2019}, or ionic liquids where the ions form immobile charged aggregates, thus leaving unlike numbers $n_\pm$ of mobile ions \cite{Agostino2018}. In the bulk one always has $\varrho=0$, yet there may be net surface charges at the sample boundaries.

Many non-equilibrium situations in physical chemistry are well described by Onsager's linear relations between fluxes and forces \cite{deGroot1962}. In the present case, this means that the ion currents are linear in the thermodynamic and electric forces, 
  \be 
  J_{\pm } = \mu_{\pm } \left(  \pm n_{\pm } eE
            -  n_{\pm }Q _{\pm }\frac{\nabla T}{T}  - k_BT \nabla n_{\pm } \right) , 
  \label{eq:6}
  \ee 
where the three contributions account for electrophoresis in an electric field $E$, thermodiffusion in a temperature gradient due to the heat of transport $Q_\pm$, and gradient diffusion with Einstein coefficient $D_\pm=k_BT\mu_\pm$. 

As a consequence of the linear-response approximation, the non-equilibrium properties arise only from the generalized forces $E$, $\nabla T$, and $\nabla n_\pm$, whereas the coefficients are evaluated at thermal equilibrium, and thus do not depend on position. Thus throughout this paper, all bulk quantities such as $T$, $n_\pm$, $Q_\pm$,... are taken as constants. The only exception occurs in Section \ref{surface} where we evaluate surface effects, which are irrelevant for the thermoelectric coefficients but merely complete the physical picture. In the case of a 1D geometry, even the generalized forces are constants, and none of the bulk properties depend on position.

\section{Open system -- thermocurrents}

Thermoelectric effects are usually defined in terms of the heat flow driven by an electric field and the charge current due to a temperature gradient, with the Peltier and Seebeck linear response coefficients.  In electronic materials, this is realized by connecting the hot and cold boundaries to electrodes. A similar situation occurs for ionic conductors which are coupled to reservoirs at different potential or temperature \cite{Nernst1889}, as illustrated in Fig. 1a. 
 
\begin{figure}[t] 
\includegraphics[width=\columnwidth]{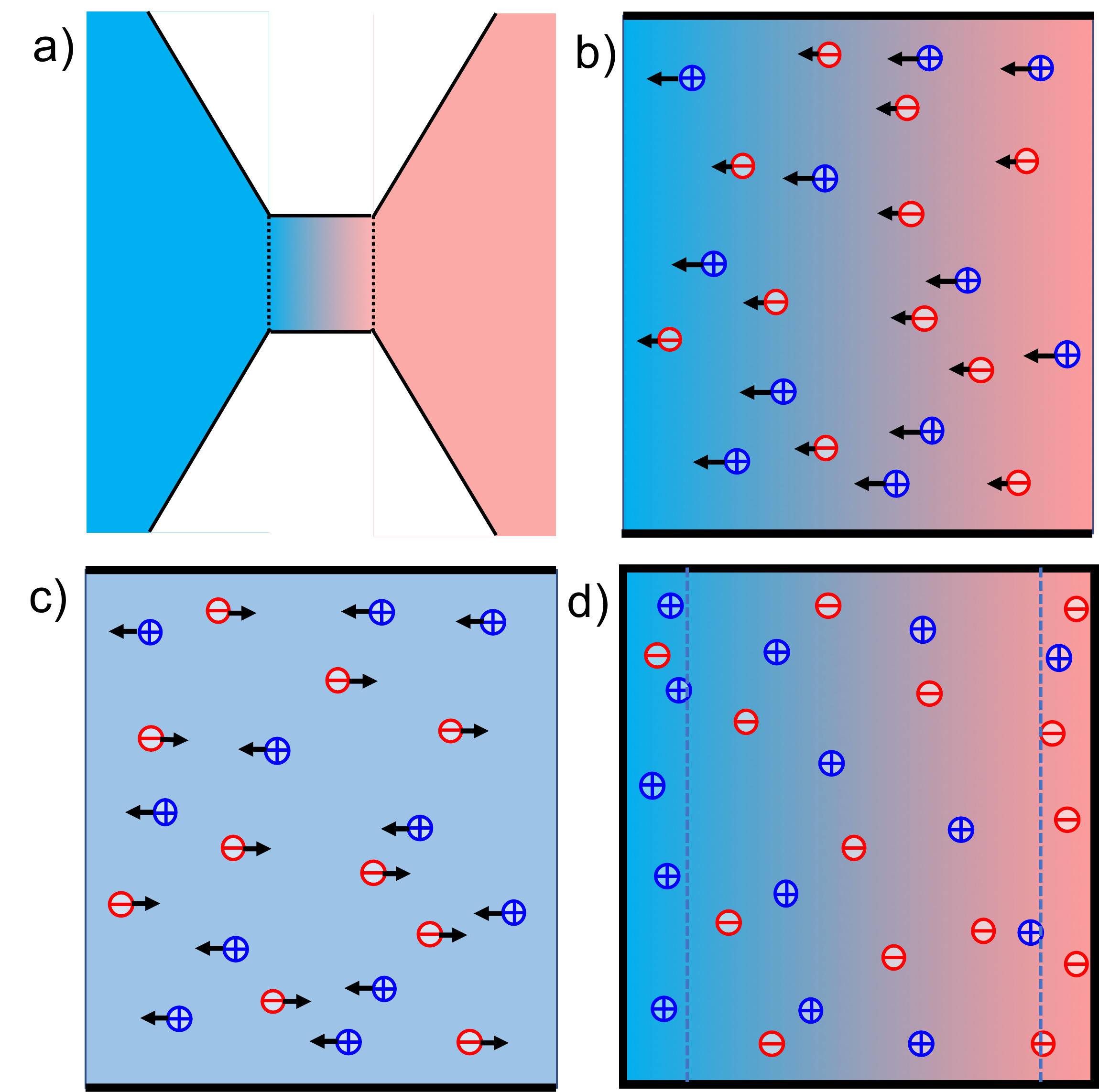}
\caption{a) Open system between two reservoirs at different temperature or different potential.  b) Seebeck effect of an open system in a thermal gradient. Due to thermodiffusion, cations and anions migrate towards the cold and carry an electric current $I_T$, resulting in a stationary current from one reservoir to the other. The picture shows the case $Q_+>Q_->0$. c) Electric conductivity and Peltier effect of an open electrolyte system in an electric field. Cations and anions move in opposite directions, inducing an electric current $I_E$ and, due to the ionic heat of transport $Q_\pm$, a heat current $\dot q_E$. d)  Steady state of a closed system, with zero ion currents $J_\pm$. There are layers of positive or negative charges within one screening length $\lambda$ from the cold and hot boundaries.}
\label{default} 
\end{figure} 

The electric conductivity $\sigma$ is defined through Ohm's law  for the current, $I_E = \sigma E$. Collecting the contributions of positive and negative carriers in $I_E=e(J_+-J_-)$, we have
 \be
  I_E = e^2 (n_+ \mu_+ +n_- \mu_-)E = ( \sigma_+   + \sigma_-)E ,
  \label{eq:12} 
  \ee
as illustrated in Fig. 1c. The relative contributions of cations and anions are expressed through Hittorf transport numbers, 
  \be 
  t_\pm = \frac{\sigma_\pm}{\sigma} = \frac{n_\pm \mu_\pm}{n_+ \mu_+ + n_- \mu_-}.
  \label{eq:14}
  \ee

Besides the charge current $I_E$, the electric field induces a heat flow: Because of their electrostatic self-energy and the interaction potential with the surrounding material, the ions carry a heat of transport $Q_{\pm}$, resulting in the heat current 
  \be
  \dot q_E =   (Q_+ n_+ \mu_+ - Q_- n_- \mu_- )eE \equiv \Pi \sigma E ,
  \ee
where the second identity defines the Peltier coefficient $\Pi$, which is readily expressed as 
  \be
  \Pi = \frac{t_+Q_+ - t_-Q_-}{e}.
  \label{eq:16}
  \ee

On the other hand, a temperature gradient $\nabla T$ gives rise to thermodiffusion of the mobile ions \cite{deGroot1962}: According to the second law, the excess enthalpy $Q_\pm$  flows towards the cold and drags the ions at velocities $-\mu_\pm Q_\pm \nabla T/T$, as illustrated in Fig. 1b. Thus the temperature gradient drives an electric current   
  \be
  I_T =  - e \left( Q_+n_+\mu_- - Q_-n_-\mu_- \right) \frac{\nabla T}{T} \equiv  - S \sigma \nabla T,
  \label{eq:18} 
  \ee
where the last identity defines the Seebeck coefficient $S$. With the above expressions for $\sigma$ and the transport numbers $t_\pm$ one finds 
  \be
  S = \frac{t_+Q_+ - t_-Q_-}{eT}.
  \label{eq:20} 
  \ee
As expected, the thermoelectric coefficients verify Onsager's reciprocal relation $S = \Pi/T$ \cite{deGroot1962}. 

In Fig. 2 we plot the Seebeck coefficient (\ref{eq:20})  as a function of the heat of transport ratio $Q_+/Q_-$, for different values of the mobility ratio $\mu_+/\mu_-$. Not surprisingly, a large cation mobility results in $S>0$, whereas highly mobile anions favor a negative Seebeck coefficient.

The above thermoelectric coefficients describe an ionic conductor sandwiched between two reservoirs at different temperature or potential. Most experiments on ionic systems, however, are done on closed systems where the charge carriers cannot enter or leave. Then the linear response for heat and electric currents is  valid only for the transient behavior after switching on the fields, or if the applied electric and temperature fields oscillate in time. For oscillatory fields the validity is restricted to sufficiently high frequency, $\omega\tau \gg 1$, where  $t\ll\tau = \lambda^2/D$ is the diffusion time of ions over one Debye screening length.  Similarly, $S$ describes the behavior after switching on the temperature gradient, for times shorter than $\tau$, which is much shorter than the time scale for the emergence of the bulk ion gradients, $D/L^2$, over the system size $L$. The transient behavior requires to solve the continuity equation $\partial_t n_\pm + \mathrm{div} J_\pm=0$ \cite{Bonetti2015,Ly2018,Janssen2019,Sehnem2020}. 

\begin{figure}[t] 
\includegraphics[width=\columnwidth]{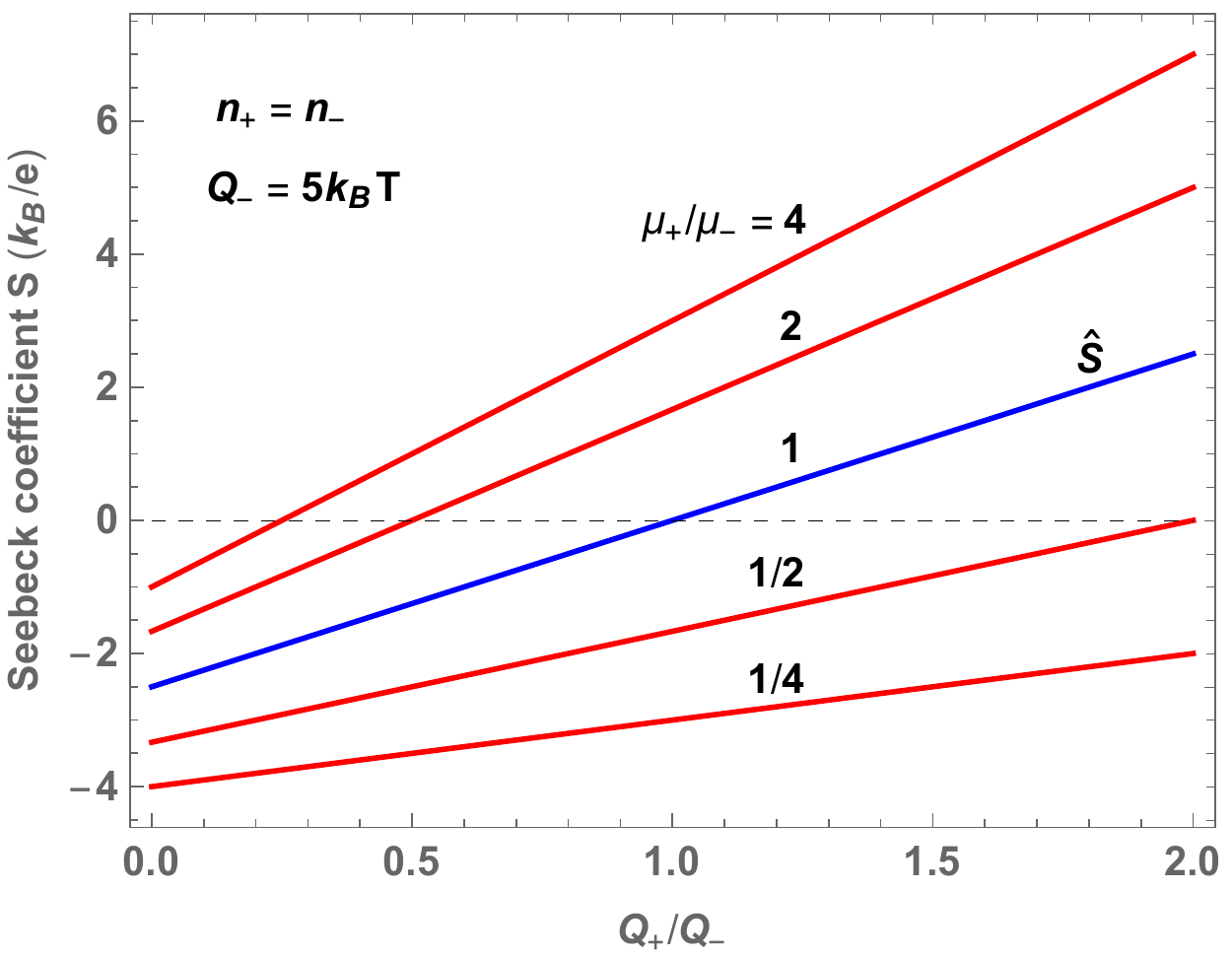}
\caption{Thermoelectric coefficients $S$ and $\hat S$ as a function of the ratio $Q_+/Q_-$. The coefficient for open systems, $S$, depends on the ratio $\mu_+/\mu_-$, whereas that for a closed systems, $\hat S$, is independent of the mobilities. Note $S=\hat S$ for $\mu_+=\mu_-$. The coefficients $S$ and $\hat S$ are given in units of $k_B/e$, for equal cation and anion concentrations and $Q_-=5k_BT$. }
\label{default}
\end{figure}

\section{Closed system -- thermopotential} 

Now we turn to a closed system which exchanges heat with the surrounding but which does not transfer charges.  After switching on the temperature gradient, the electric current $I_T$ accumulates charges at the hot and cold boundaries,  which in turn give rise to an electric field. After a transient time, the system attains a steady state with constant surface charge and zero ion currents (Fig. 1d), which is characterized by the thermoelectric field
  \be
  E = \hat S \nabla T,  
\label{eq:21}
  \ee
or by the corresponding thermopotential $-\hat S (T_H-T_C)$ between hot and cold boundaries. Previous works on ionic thermoelectrics often assume, more or less explicitly, that the steady state is characterized by the Seebeck coefficient (\ref{eq:20}), that is, $\hat S=S$.

\subsection{Steady state} 

We start with the fundamental equations describing the steady state of a system that is open with respect to heat flow but closed for charge carriers, as shown in Fig. 1d. First, the currents of both cations and anions vanish, 
  \be
  J_\pm = 0. 
   \label{eq:22}
  \ee
 Second, the electric field is related to the charge density by Gauss's law  
  \be
  \varrho =\varepsilon\nabla\cdot E. 
  \label{eq:25}
  \ee
Third, supposing a one-dimensional geometry as in Fig. 1d, the electric field vanishes at the solid boundary, 
  \be
  E|_B = 0.
  \label{eq:23}
  \ee

In the following we first evaluate the bulk thermoelectric field (\ref{eq:21}). In a second step we study the electrostatic properties of the surface layers shown in Fig. 1d, in order to satisfy the boundary condition (\ref{eq:23})

\subsection{Thermoelectric field} 

Here we derive the coefficient $\hat S$ defined in  (\ref{eq:21}).  Inserting the currents (\ref{eq:6}) in $J_+-J_- = 0$, solving for the thermoelectric field $E$,  and using the Seebeck coefficient (\ref{eq:20}), we obtain  
  \be
  E = S \nabla T +   \frac{k_BT}{e} \frac{\mu_+ \nabla n_+ - \mu_- \nabla n_-}{n_+ \mu_+ + n_- \mu_-}.    
  \label{eq:28}
  \ee
In order to evaluate the latter term, we note that the bulk charge density vanishes, $\varrho=0$. In view of (\ref{eq:4}) this means that the unperturbed bulk concentrations satisfy $e(n_+^0-n_-^0)+\varrho_f=0$, and that the concentration gradients of cations and anions are identical,
  \be
  \nabla n_+ = \nabla n_- .  
  \label{eq:29}
  \ee
Solving the equation $n_-J_+ + n_+ J_- = 0$ for this gradient we obtain
  \be
  \nabla n_\pm = - \frac{Q_+ + Q_-}{k_BT} \frac{n_+ n_-}{n_+ + n_-}  \frac{\nabla T}{T}.
  \label{eq:29a}
  \ee
Insertion in (\ref{eq:21}) finally gives the thermoelectric coefficient  
  \be
  \hat S = S + \frac{Q_++Q_-}{eT}
                     \frac{\mu_+ - \mu_-} {n_+ \mu_+ + n_- \mu_-}\frac{n_+ n_-}{n_+ + n_-}.    
  \label{eq:30}
  \ee
Thus the thermoelectric coefficients of closed and open systems, $\hat S$ and $S$, differ by a term which is proportional to the mobility contrast $\mu_+-\mu_-$ of cations and anions.

After inserting (\ref{eq:20}) and rearranging the terms, this expression significantly simplifies,
  \be
  \hat S = \frac{\hat t_+Q_+ - \hat t_-Q_-}{eT},
  \label{eq:24}
  \ee
with the weight factors 
  \be 
  \hat t_\pm = \frac{n_\pm }{n_+  + n_-}. 
  \label{eq:26}
  \ee
As a striking feature, we find that the weight factors $\hat t_\pm$ and thus the coeffcient $\hat S$,  are independent of the ionic mobilities, contrary to the Seebeck effect $S$ defined through the thermocurrent. 

For identical mobilities ($\mu_+=\mu_-$) we have $\hat S = S$, whereas in the general case these coefficients differ significantly, and may even take opposite signs. This is illustrated by Fig. 2, where we compare $\hat S$ and $S$ as a function of the heat of transport ratio $Q_+/Q_-$. The curves for different mobility ratio $\mu_+/\mu_-$ highlight the fundamentally different behavior expected for open and closed ionic systems.

\subsection{Surface effects}\label{surface}

For the thermoelectric coefficients calculated above, we have not used the boundary condition (\ref{eq:23}) for the electric field. In order to complete the physical picture and to account for the thermocharge at the hot and cold boundaries, we now derive the surface electric field, required to satisfy (\ref{eq:23}). For a 1D geometry with constant $\nabla T$, we solve the Poisson-Boltzmann equation $\nabla^2\psi+\varrho/\varepsilon=0$ in Debye-H\"uckel approximation, and obtain the homogeneous potential $\psi_h=\psi_0 \sinh(x/\lambda)$ and field $E_h = - \partial_x \psi_h$, with the Debye screening length $\lambda = \sqrt{e^2(n_++n_-)/\varepsilon k_BT}$. 

Adjusting the prefactor $\psi_0$ in view of (\ref{eq:21}) and (\ref{eq:23}), one readily finds the total electric field
  \be
    E(x) = \hat S \nabla T\left(1- \frac{\cosh(x/\lambda)}{\cosh(L/2\lambda)} \right),
  \ee
which agrees with both the bulk value (\ref{eq:21}) and the boundary condition (\ref{eq:23}). In experimental situations, the system size $L$ is by at least several orders of magnitudes larger than $\lambda$.

The charge density is obtained from Gauss's law (\ref{eq:25}), 
  \be
  \varrho(x) = - \frac{\varepsilon \hat S \nabla T}{\lambda} \frac{\sinh(x/\lambda)}{\cosh(L/2\lambda)},
  \ee
which, for $\hat S>0$, is positive at the cold surface ($x=-L/2$) and negative at the hot one ($x=L/2$), as illustrated in Fig. 1d. Beyond a few Debye lengths from the boundaries, the electric field takes the constant value and the charge density vanishes, as anticipated in (\ref{eq:29}).

\section{Mixed electrolytes}

For the sake of simplicity we have so far considered monovalent binary electrolyte solutions. In order to account for more complex systems, containing mixtures of different salts or acids \cite{Putnam2005,Vigolo2010,Wuerger2008,Eslahian2014,Sehnem2020}, or multivalent ions \cite{Sehnem2020}, we give the general expression, with several cations and anions species $i$, of  concentration $n_i$ and valency $z_i$. 

The Seebeck coefficient of an open system is readily generalized in terms of the partial conductivity $\sigma_i = z_i^2e^2n_i\mu_i$, resulting in
  \be
   S = \frac{1}{eT} \frac{ \sum_i z_i n_i \mu_i Q_i}{ \sum_i z_i^2 n_i \mu_i} 
  \label{eq:40}
  \ee 
and Hittorf transport numbers $t_i=\sigma_i/\sigma$.

Now we turn to the steady-state of a closed system and generalize the coefficient $\hat S$. Inserting the zero-current and zero-charge conditions, 
   \be
   J_i = 0, \;\;\;  \sum_iz_i \nabla n_i = 0 ,
  \label{eq:41}
   \ee 
in the charge current $I=e \sum_i  z_i J_i$, eliminates the ion gradients. Solving for the electric field $E$ one readily obtains 
  \be
  \hat S = \frac{1}{eT} \frac{ \sum_i z_i n_i Q_i}{ \sum_i z_i^2 n_i} .
  \label{eq:42}
  \ee 
This form generalizes the expressions used previously for the mixed electrolyte NaCl$_x$OH$_{1-x}$ where $z_i=\pm1$ \cite{Wuerger2008,Vigolo2010,Eslahian2014}. A more complex form occurs if one of the species is a macroion, for example a charged polymer or a colloidal particle, where the thermally driven velocity cannot be expressed by the product of Peltier heat $Q$ and mobility $\mu$, and where both the numerator and the denominator in (\ref{eq:42}) depend on the ratio of phoretic and diffusive mobilities \cite{Majee2012}.

\section{Discussion}

As the main result of this paper, we found that the thermoelectric properties of ionic conductors depend crucially on the boundary conditions: For an open system, the Seebeck coefficient  is defined through the thermally induced current $I_T$, where the heat of transport of positive and negative carriers is weighted with Hittorf's transport numbers $t_\pm$. The resulting coefficient (\ref{eq:20}) depends on the mobilities, similar to what is known for electronic systems \cite{Fritsche1971}.

For a closed system, on the contrary, the coefficient $\hat S$ is defined through the macroscopic thermoelectric field, or the potential difference between the hot and cold boundaries,
  \be
  V_H-V_C= - \hat S(T_H - T_C). 
  \label{eq:36}
  \ee
The thermopower $\hat S$, given in Eq. (\ref{eq:24}), differs significantly from the Seebeck coefficient $S$ and in particular does not depend on the ion mobilities, as is obvious when comparing the weight factors $\hat t_\pm$ and $ t_\pm$.  Recent experiments on various ionic thermoelectric materials  \cite{Zhao2016,Kim2016,Wang2017,Li2019,Zhao2019,Han2020} do not generate currents but a thermopotential, and thus are described by the coefficient $\hat S$.

  \begin{figure} 
  \includegraphics[width=\columnwidth]{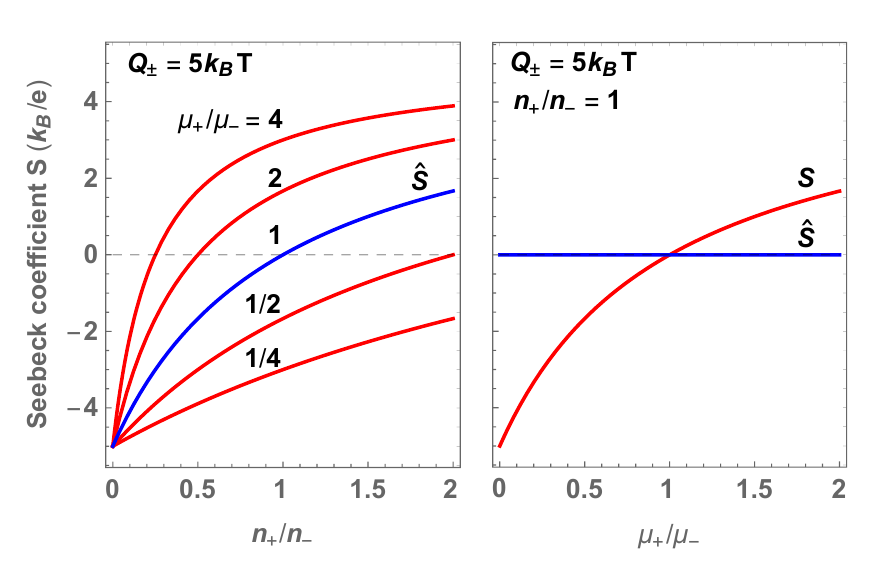}
  \caption{Thermoelectric coefficients $S$ and $\hat S$ as a function of the concentration ratio $n_+/n_-$ and the mobility ratio $\mu_+/\mu_-$. For unlike mobilities, $S$ (red) and $\hat S$ (blue) differ significantly, whereas $S=\hat S$ for $\mu_+=\mu_-$. The coefficients $S$ and $\hat S$ are given in units of $k_B/e$, with equal cation and anion heat of transport $Q_\pm=5k_BT$. }
  \label{default}
  \end{figure}  
 
\begin{table}[b]
\caption{Mobility data of salts, hydrochloric acid, and ionic liquids consisting of the cation EMIM$^+$ (1-ethyl-3-methylimidazolium) and the anions TFSI$^-$ (bis(trifluoro-methylsulfonyl)imide) or OAc$^-$ (acetate). Mobilities $\mu_\pm=D_\pm/k_B T$ are calculalted from measured diffusion coefficients $D_\pm$ \cite{Agar1989,Agostino2018}.}
\begin{tabular}{|l|c|c|c|}
\hline
     &  $\mu_+$ ($\mathrm{s/\mu g}$)  & $\mu_-$ ($\mathrm{s/\mu g}$) & $\mu_+/\mu_-$  \\   \hline      
    LiCl \cite{Agar1989}    & 25.1   &  49.5    & 0.51              \\ \hline   
    NaCl \cite{Agar1989}    & 32.5   &  49.5    & 0.66             \\ \hline   
    HCl \cite{Agar1989}    & 227  &  49.5    & 4.59             \\ \hline 
    EMIM-TFSI\cite{Agostino2018}    & 5.75 &  3.39  & 1.70 \\ \hline 
    EMIM-OAc \cite{Agostino2018}   & 2.63  & 2.57 &   1.02   \\ \hline        
    EMIM-OAc-glucose \cite{Agostino2018}   & 0.64 & 0.25 &  2.6  \\ \hline         
 \end{tabular}
\linebreak
\end{table}

\subsection{Experimental relevance}

The difference between the coefficients $S$ and $\hat S$ may be quite significant, and is best illustrated in terms of the ratios of  heats of transport $Q_+/Q_-$, concentrations $n_+/n_-$, and mobilities $\mu_+/\mu_-$. Fig. 2 compares $S$ and and $\hat S$ as a function of the heat of transport ratio for different mobility contrasts.  In Fig. 3a we plot the thermoelectric coefficients as a function of the concentration contrast, for different mobilities but equal heats of transport; this case is realized, for example, by the ionic liquid EMIM$^+$TFSI$^-$, where the cations have a larger mobility but there are ten times more mobile anions \cite{Agostino2018}. Finally, Fig. 3b shows these coefficients as a function of the mobility contrast, at equal heat of transport of cations and anions. 

In order to illustrate the difference of the coefficients $S$ and $\hat S$, we give in Table 1 the mobility contrast of several electrolyte solutions and ionic liquids; because of their molecular mass contrast, even larger values are expected for polymer electrolytes like Na$^+$PEG-OH$^-$ \cite{Zhao2016,Li2019}. Note that adding a component, such as sugar to the ionic liquid EMIM-OAc, may modify the mobility ratio. Together with the curves of Figs. 2 and 3, the numbers of Table 1 provide evidence that for a given ionic system, the boundary conditions strongly affect the thermoelectric properties, and in many cases result in opposite signs of $S$ and $\hat S$. Note that in dense electrolytes, such as ionic liquids, correlation effects may become important \cite{Demery2016} and modify the heat of transport $Q_\pm$.

\subsection{The salt gradient as companion field}

It turns instructive to rewrite the charge current $I=e(J_+-J_-)$ in terms of the above coefficients,
  \be
  I = \sigma E - S\sigma \nabla T - (\mu_+ - \mu_-) e k_BT \nabla n_\pm.
  \label{eq:38}
  \ee
For an open system in a temperature gradient, one has $E=0$ and $\nabla n =0$, and one readily recovers the definition of the Seebeck coefficient $S$. 

In a closed system, the boundary condition  imposes $I=0$. In the case of a single conducting species, Eq. (\ref{eq:29}) implies $\nabla n_\pm=0$. Then the thermoelectric field $E=S\nabla T$ is given by the Seebeck coefficient, and generates the current $I_E$  which cancels the thermodiffusion current $I_T$. 
If there are two carrier species, however, their currents must vanish separately, $J_\pm=0$, thus giving rise to a finite concentration gradient (\ref{eq:29a}), which induces gradient diffusion. In other words, Eq. (\ref{eq:38}) consists of a thermocurrent $I_T$, the electric current $I_E$ induced by the Seebeck field $E$, and a gradient diffusion current $I_n$, resulting in both $I=0$ and $J_\pm=0$. 

Both the thermoelectric field $E$ and the concentration gradient $\nabla n_\pm$ may be viewed as companion fields of the temperature gradient, which are imposed by the boundary conditions of a closed system. The former is described by $\hat S$, whereas the latter may be expressed through the salt Soret coefficient $S_T$, which is defined through $\nabla \ln(n_++n_-) + S_T \nabla T=0$  \cite{Wiegand2004} and reads explicitly
  \be
  S_T =  \frac{Q_+ + Q_-}{k_BT^2} \frac{2n_+ n_-}{(n_+ + n_-)^2} . 
  \ee
Then the parameter $\hat S$ consists of two contributions,
  \be
  \hat S = S - \frac{k_BT}{2e} \frac{(\mu_+ - \mu_-)(n_+ + n_-)} {n_+ \mu_+ + n_- \mu_-} S_T,
  \label{eq:40}
  \ee
where the first one is the usual Seebeck coefficient, and the second one arises from the Soret effect of the electrolyte. 
Such ``companion fields'' have been reported for various examples of thermally driven motion: Colloidal thermophoresis has been shown to be often dominated by gradients of added polymer \cite{Jiang2009} or salt \cite{Eslahian2014}, whereas self-diffusiophoresis of hot Janus particles in near-critical binary liquids is driven by the non-uniform composition in the particle's vicinity \cite{Wuerger2015}.

\subsection{Comparison with electronic conductors}

We conclude by comparing the present findings to what is known for semiconductors and metals. The Peltier heat of transport of electrons and holes is given by the energy with respect to the Fermi level, $E-E_F$. Integrating separately over valence and conduction bands $V$ and $C$, one has 
  \be
  t_\pm Q_\pm = \mp \int_{V,C} dE \frac{\sigma_E}{\sigma} (E - E_F) , 
  \ee
with the conductivity $\sigma=\int dE \sigma_E$. Then the Seebeck coefficient given by Fritsche \cite{Fritsche1971}, which generalizes Mott's formula \cite{Cutler1969} to non-uniform conductivity, is identical to our Eq. (\ref{eq:20}). 

Yet a difference arises for closed systems. Unlike ionic concentrations, the numbers of electrons and holes are not conserved individually: Thermal excitation and recombination permanently create and annihilate  carriers.  Because of these ``chemical reactions", the individual currents $J_\pm$ need not to vanish in electronic conductors, contrary to (\ref{eq:12}) for ions. Since the carrier concentration is imposed by the local chemical potential, there is no additional diffusion current. As a consequence, the concentration gradient terms are missing in Eqs. (\ref{eq:28}) and (\ref{eq:38}), and the thermoelectric field is given by the same coefficient $S$ as the thermocurrent in an open system. This leads us to the conclusion that the particular properties of ionic conductors stem from the presence of chemically inert cations and anions. 

The author thanks X. Crispin and S. Nakamae for stimulating discussions, and M. Janssen for helpful remarks on a first draft of the manuscirpt. This project was supported by the French National Research Agency through grant ANR-19-CE30-0012-01 and by the European Research Council (ERC) through grant n$^\mathrm{o}$ 772725. 
 
\bibliography{../../-Archive/literature}

\end{document}